\documentclass[aps, twocolumn, prd, preprintnumbers, amsmath, amssymb, amsfonts, superscriptaddress, nofootinbib]{revtex4-1}
\pdfoutput=1

\usepackage[linktocpage,pagebackref=false,hidelinks]{hyperref}

\usepackage{setspace,latexsym}
\usepackage{color}
\usepackage{epsfig}
\usepackage{graphicx}
\usepackage{slashed}
\usepackage[export]{adjustbox}

\newcommand{\beq}{\begin{equation}}
\newcommand{\eeq}{\end{equation}}
\newcommand{\bea}{\begin{eqnarray}}
\newcommand{\eea}{\end{eqnarray}}
\newcommand{\nn}{\nonumber}

\newcommand{\mueV}{\mathrm{\mu eV}}
\newcommand{\meV}{\mathrm{meV}}
\newcommand{\GeV}{\mathrm{GeV}}
\newcommand{\MeV}{\mathrm{MeV}}
\newcommand{\TeV}{\mathrm{TeV}}

\newcommand{\SU}{\mathrm{SU}}
\newcommand{\U}{\mathrm{U}}

\newcommand{\C}{\mathrm{c}}
\newcommand{\I}{\mathrm{i}}

\newcommand{\dd}{\mathrm{d}}

\newcommand{\PS}{{\phantom*}}
\newcommand{\PP}{{\phantom\prime}}
\newcommand{\PD}{{\phantom\dag}}

\newcommand{\SM}{\mathrm{SM}}
\newcommand{\EW}{\mathrm{EW}}
\newcommand{\PQ}{\mathrm{PQ}}
\newcommand{\LD}{\mathrm{LD}}
\newcommand{\kd}{\mathrm{kd}}
\newcommand{\eff}{\mathrm{eff}}
\newcommand{\Mpl}{M_\mathrm{Pl}}
\newcommand{\UV}{\mathrm{UV}}
\newcommand{\trap}{\mathrm{trap}}
\newcommand{\QCD}{\mathrm{QCD}}
\newcommand{\DM}{\mathrm{DM}}

\begin{document} 

\title{Composite neutrinos and the QCD axion:\\
baryogenesis, dark matter, small Dirac neutrino masses, and vanishing neutron EDM}

\author{Sabyasachi Chakraborty}
\email{schakraborty5@fsu.edu}
\affiliation{Department of Physics, Florida State University, Tallahassee, FL 32306, USA}

\author{Tae Hyun Jung}
\email{thjung0720@gmail.com}
\affiliation{Department of Physics, Florida State University, Tallahassee, FL 32306, USA}

\author{Takemichi Okui}
\email{tokui@fsu.edu}
\affiliation{Department of Physics, Florida State University, Tallahassee, FL 32306, USA}
\affiliation{High Energy Accelerator Research Organization (KEK), Tsukuba 305-0801, Japan}

\preprint{KEK-TH-2341}

\begin{abstract}
We consider a chiral gauge theory from which light composite Dirac neutrinos dynamically emerge, augmented by a QCD axion to solve the strong CP problem. We show that an interplay between the composite sector and the axion can also simultaneously lead to successful baryogenesis and generate a correct dark matter abundance via ``co-genesis'' without contradicting present constraints. We predict $\Delta N_{\rm eff} \geq 0.14$ and $m_a$ in the range $O(10)\>{\rm \mu eV}$--$O(10)\>{\rm meV}$, which can be firmly tested by upcoming CMB experiments and axion searches.
\end{abstract}

\maketitle

\section{Introduction}
Why are neutrinos so light? Why does lepton number appear to be conserved so well?
These questions have a beautiful answer for Majorana neutrinos---gauge invariance simply dictates that neutrino masses and associated lepton number violation must originate from an irrelevant interaction suppressed by a large mass scale.
But what if neutrinos are Dirac?
Do we have a similarly compelling theoretical plot where small neutrino masses and associated lepton number conservation simply emerge from gauge invariance?  

Such was provided a long ago by Arkani-Hamed and Grossman~\cite{ArkaniHamed:1998pf}.
They augmented the SM by a new confining chiral gauge theory whose low energy spectrum below confinement just consists of \emph{massless} ``baryons'' $\nu^\C$. 
For example, imagine $\nu^\C \sim \Psi\Psi\Psi / \Lambda_\C^3$ schematically, where $\Psi$ denotes the ``quarks'' of the new chiral gauge sector and $\Lambda_\C$ its confinement scale.
Then, the leading gauge invariant interaction between $\nu^\C$ and SM fields is given by a highly irrelevant operator $H\ell\Psi\Psi\Psi / M^3$ suppressed by a high scale $M$. 
Below $\Lambda_\C$, this operator becomes a Yukawa interaction $y_\nu H\ell\nu^\C$ with $y_\nu \sim (\Lambda_\C/M)^3$,
so a mild hierarchy between $\Lambda_\C$ and $M$ can give us a tiny $y_\nu$.
This operator accidentally conserves lepton number with $\nu^\C$ carrying lepton number $-1$.
Gauge invariance dictates that the leading violation of lepton number conservation comes from $(\Psi\Psi\Psi)^2 / M^5 \sim \Lambda_\C(\Lambda_\C/M)^5 \nu^\C\nu^\C$, an even more irrelevant operator than $H\ell\Psi\Psi\Psi$.
Thus, $\nu^\C$ and the SM neutrino $\nu$ together acquire a naturally small Dirac mass,
with the Majorana mass for $\nu^\C$ and associated lepton number violation being automatically negligible.
Light Dirac neutrinos and lepton number conservation thus emerge from gauge invariance.%
\footnote{This mechanism has an obvious RS dual~\cite{Gherghetta:2003hf} if we replace the asymptotically-free gauge dynamics by a strongly coupled CFT\@.}

In this paper, we show that just such composite sector for explaining small Dirac neutrino masses augmented by a QCD axion to solve the strong CP problem (i.e., make the neutron EDM vanish) can also simultaneously provide a correct DM abundance and successful baryogenesis, over a large viable parameter range. 
The first two of these four problems are, of course, solved by construction.
However, DM and baryon asymmetry are generated through an intimate interplay between the axion and composite sectors via the ``co-genesis'' mechanism~\cite{Co:2019wyp, Co:2019jts, Co:2020xlh}, 
and it is this interplay that permits a large range of parameters in which all four problems are simultaneously solved.

\section{What goes in:\\ Neutrino masses and strong CP}
As described above, our theoretical inputs are a composite sector for light Dirac neutrinos and a QCD axion for the strong CP problem. 
For concreteness we study the following composite sector,
but our framework itself is more general and not intrinsically tied to this specific example.  

Consider an $\SU(6)$ gauge theory with three left-handed Weyl fermions, $\psi_i$ ($i=1,2$) and $\chi$, 
where $\psi_i$ and $\chi$ are in the $\mathbf{\bar{6}}$ and $\mathbf{15}$ representations of $\SU(6)$, respectively.
No masses and interactions with mass dimension $\leq 4$ are allowed by the $\SU(6)$ gauge invariance beside the gauge interaction itself.
Consequently, the composite sector Lagrangian \emph{accidentally} possesses $\SU(2)_\psi$$\times$$\U(1)_L$ global symmetries shown in Table~\ref{tab:charges}, which are anomaly free under the $\SU(6)$ gauge interaction and are thus true emergent symmetries of the theory.

Now, what are the light degrees of freedom of the composite sector and how can they interact with SM particles?
As discussed in~\cite{Dimopoulos:1980hn}, 
we have compelling arguments, though not a proof, that strongly suggest that this gauge theory undergo confinement without chiral symmetry breaking, where the sole degrees of freedom below the confinement scale are three \emph{massless} spin-$1/2$ ``baryons'' interpolated by the composite operators $\psi_{\{i} \psi_{j\}} \chi$, 
where $\{i\cdots j\}$ indicates symmetrization of the $\SU(2)_\psi$ doublet indices.%
\footnote{ The conjecture of~\cite{Dimopoulos:1980hn} is that an $\SU(N)$ gauge theory with $N-4$ Weyl fermions in the $\mathbf{\overline{N}}$ representation and one Weyl fermion in the $\mathbf{N(N-1)/2}$ representation should have $(N-3)(N-4)/2$ massless ``baryons'' with the $\SU(N-4)$ flavor symmetry unbroken.
The $N=6$ case gives exactly 3 $\nu^\C$'s. 
A different conjecture was proposed in~\cite{Csaki:2021xhi} that the $\SU(N-4)$ should be spontaneously broken to $\mathrm{Sp}(N-4)$ ($\mathrm{Sp}(N-5)$) for $N = \text{even}$ (odd).
Amusingly, for our case of $N=6$, the two conjectures agree because $\mathrm{Sp}(2) = \SU(2)$! 
Discrete symmetries are also consistent with our case as analyzed carefully in~\cite{Smith:2021vbf}.}
The low energy spectrum of the theory, therefore, is given by these three massless Weyl fermions + SM degrees of freedom.
The leading gauge-invariant interaction between the massless ``baryons'' and SM fields appears at dimension-7 and is given by
\bea
\frac{1}{M^3} H\ell\psi\psi\chi 
\label{eq:Hlqqq}
\eea
with some high scale $M$, where $H$ and $\ell$ are the SM Higgs and lepton doublets.
Hence, following~\cite{ArkaniHamed:1998pf},
we identify the three ``baryons'' with three gauge-singlet fermions for Dirac neutrinos:
\bea
\nu^\C \sim  \frac{(4\pi)^2}{N_\C \Lambda_\C^3} \psi\psi\chi 
\,,
\eea
where $N_\C = 6$ and we have estimated the coefficient by assuming the 't Hooft coupling, $N_\C g_6^2/(4\pi)^2$, is $O(1)$ at scales around $\Lambda_\C$.
Tiny Yukawa couplings for Dirac neutrinos (e.g., $y_\nu \simeq 3 \times 10^{-13}$ for $m_\nu \simeq 0.05~\mathrm{eV}$) are then generated below the $\SU(6)$ confinement scale $\Lambda_\C$ as:
\bea
\text{operator~(\ref{eq:Hlqqq})} \longrightarrow y_\nu H\ell\nu^\C
\,,\quad
y_\nu \sim \frac{N_\C}{(4\pi)^2} \!\left( \frac{\Lambda_\C}{M} \right)^{\!\!3}
\,.
\label{eq:numass}
\eea
Thanks to the third power of $\Lambda_\C / M$, the extreme smallness of $y_\nu$ is reduced to
a moderate hierarchy between $\Lambda_\C$ and $M$ (roughly $\Lambda_\C / M \sim 10^{-4}$)~\cite{ArkaniHamed:1998pf}.
The neutrino masses can have Majorana components coming from the operator $(\psi\psi\chi)^2$, but this has dimension 9 so the Majorana components are negligible compared to the Dirac components. 
The neutrinos thus \emph{have} to be Dirac~\cite{ArkaniHamed:1998pf}.

\begin{table}[t]
\begin{tabular}{c||c|cc}
                               & ~gauge~            & \multicolumn{2}{c}{global}  \\ \hline
                               & $\SU(6)$           & ~$\SU(2)_\psi$~ & $\U(1)_L$ \\ \hline\hline
$\psi$                         & $\mathbf{\bar{6}}$ & $\mathbf{2}$  & $-2/3$      \\
$\chi$                         & $\mathbf{15}$      & $\mathbf{1}$  & $1/3$       \\ \hline  
$\nu^\C \propto \psi\psi\chi~$ & $\mathbf{1}$       & $\mathbf{3}$  & $-1$        
\end{tabular}
\caption{Symmetries in the composite sector. 
$\SU(6)$ is gauged, 
while $\SU(2)_\psi$ and $\U(1)_L$ emerge as accidental global symmetries of the composite sector,
both free of anomalies under $\SU(6)$.
The Weyl fermions $\psi$ and $\chi$ are elementary, while $\nu^\C$ is a massless composite ``baryon'' from $\SU(6)$ confinement, carrying the quantum numbers of $\psi\psi\chi$.}
\label{tab:charges}
\end{table}

The accidental $\U(1)_L$ of the composite sector is identified with the accidental $\U(1)$ lepton number of the SM by the operator~(\ref{eq:Hlqqq}), hence its name.
While $\U(1)_L$ is anomaly free under $\SU(6)$, it is now anomalous under $\SU(2)_W$$\times$$\U(1)_Y$.
$\U(1)_{B-L}$ is an accidental, anomaly-free global symmetry of the entire SM+composite theory.

$\SU(2)_\psi$ is explicitly broken by charged lepton Yukawa couplings as well as the operator~(\ref{eq:Hlqqq}) since gauge invariance permits an arbitrary 3$\times$3 matrix for its coefficients.
Nevertheless, thanks to the hierarchy $\Lambda_\C \ll M$ and a high dimension of the operator~(\ref{eq:Hlqqq}), properties of the composite sector, such as the existence of three massless fermions, should not be changed.
 
Our final ingredient is an axion.
To solve the strong CP problem, the Lagrangian must include $a \,\mathrm{tr}[G_{\mu\nu}\widetilde{G}^{\mu\nu}]$, where $a$ denotes the axion, $G_{\mu\nu}$ the gluon field strength tensor, and $\widetilde{G}^{\mu\nu} \equiv \epsilon^{\mu\nu\rho\sigma} G_{\rho\sigma} / 2$:
\bea
\mathcal{L} \supset c_3 \frac{\alpha_3}{4\pi} \frac{a}{f_a} \mathrm{tr}[G_{\mu\nu}\widetilde{G}^{\mu\nu}]
\,.
\label{eq:aGG}
\eea
The axion may have more couplings than this but we will see this minimal case with only $c_3$ can already provide DM and baryogenesis via its interplay with the composite sector.

\section{What comes out:\\ Baryogenesis and DM,\\ and more}
While our minimal model above structurally ensures that the neutrinos are Dirac and light and that the strong CP problem is solved,
the neutrino and axion sectors appear just two independent modules to solve two separate problems. 
Intriguingly, it turns out that, without any further addition, the two sectors together can also generate correct baryon asymmetry and DM abundance via the ``co-genesis'' mechanism of~\cite{Co:2019wyp, Co:2019jts, Co:2020xlh}, while still preserving the successes in neutrino masses and strong CP\@.

\subsection{Baryogenesis}
From a low energy perspective around the electroweak phase transition temperature, 
our baryogenesis is a leptogenesis in that the baryon asymmetry at the electroweak scale
originates from the $B-L$ asymmetry 
generated at some higher temperature $T_\LD$:
\bea
\eta_B^\PS \equiv \frac{\langle B \rangle}{s} \biggr|_{T_\EW}
\simeq 
\frac{28}{79} \frac{\langle B-L_\SM \rangle}{s} \biggr|_{T_\LD} 
\,,\label{eq:leptogenesis} 
\eea
where $s$ is the entropy density, $T_\EW \sim 130\>\GeV$ is the electroweak sphaleron decoupling temperature~\cite{DOnofrio:2014rug},
and $L_\SM$ ($L_\C$) is the contribution from the SM (composite) sector to the total lepton number, $L = L_\SM + L_\C$.

Unlike the usual leptogenesis from heavy right-handed neutrino decays, however, 
our nonzero value of $\langle B-L_\SM \rangle |_{T_\LD}$ is produced via the co-genesis mechanism~\cite{Co:2019wyp, Co:2020xlh}, which essentially proceeds as follows (see the Supplemental Material 
 for details). 
At sufficiently high temperatures where the axion potential is negligible, 
PQ charge conservation implies 
\bea
f_a^2 \dot\theta + \sum_P c_P n_P = f_a^2 \dot\theta_\I 
\,,\label{eq:PQ-charge-relation}
\eea
where $\theta \equiv a / f_a$ with $\dot{~} \equiv \frac{\dd}{\dd t}$, while $c_P$ is the PQ charge of particle species $P$, and $n_P$ the number density of $P$ \emph{minus} that of anti-$P$.
By assumption, at some ``initial'' temperature $T_\I$ we have $n_P(T_\I) = 0$ for all $P$ and a classical axion field configuration with $\dot\theta(T_\I) = \dot\theta_\I \neq 0$ (see~\cite{Co:2019wyp, Co:2019jts} for a possible origin of $\dot\theta_\I$ and its implications for the axion quality problem).

Scatterings among particles distribute PQ charge to the $n_P$'s such that the free energy is minimized subject to constraints from all \emph{effective} conservation laws.
In particular, the constraints include $\langle B-L \rangle = 0$ with $L = L_\SM + L_\C$.
However, at temperatures below what we call the \emph{lepton-number decoupling (LD)} temperature $T_\LD$, 
the processes shuffling lepton numbers between $L_\SM$ and $L_\C$ go out of equilibrium, thus \emph{effectively} conserving $B - L_\SM$ and $L_\C$ separately.
Minimizing the free energy under all such constraints, we get
\bea
\frac{\langle B-L_\SM \rangle}{s} \biggr|_\LD \simeq c_\LD^\PS \, \frac{T_\LD^2}{f_a^2} Y_\theta
\,,\label{eq:B-LSM}  
\eea
where $Y_\theta \equiv f_a^2 \dot\theta / s$ and $c_\LD$ is a linear combination of the $c_P$'s.
As shown in the Supplemental Material, 
 $Y_\theta$ is conserved at temperatures well below $f_a$ but still sufficiently high that the axion potential is negligible.

Most importantly, $c_\LD$ contains $c_3$. 
As shown in the Supplemental Material, 
we have $c_\LD \simeq -0.04 c_3 +\ \cdots$ or $\simeq -0.03 c_3 + \cdots$ depending on whether electron-Yukawa-mediated processes are in equilibrium or not, respectively, and the ellipses denote $c_P$'s other than $c_3$.
Hence, even the most minimal axion coupling~\eqref{eq:aGG} for the strong CP problem will work also for baryogenesis.

This is a good place to pause and compare our scenario with the minimal co-genesis with just SM + axion discussed in~\cite{Co:2019wyp, Co:2020xlh}.
In~\cite{Co:2019wyp, Co:2020xlh}, co-genesis directly performs baryogenesis (as opposed to leptogenesis), which gives $\eta_B \propto T_\EW^2$. 
In contrast, from combining~\eqref{eq:leptogenesis} and~\eqref{eq:B-LSM}, we have $\eta_B \propto T_\LD^2$,
so our baryon asymmetry is enhanced relatively by a factor of $(T_\LD / T_\EW)^2$. 
This enhancement allows us to have co-genesis \emph{and} still solve the strong CP problem by the QCD axion, while in~\cite{Co:2019wyp, Co:2020xlh} the region compatible with co-genesis and current axion bounds turns out to contradict the QCD axion relation $m_a f_a \sim m_\pi f_\pi$ 
(see Refs.~\cite{Co:2019wyp, Co:2020jtv, Harigaya:2021txz} for other proposals to enable co-genesis with the QCD axion).

The estimation of $T_\LD$ turns out to be tricky.   
One might attempt to estimate $T_\LD$ as follows.
The rate of conversion between $L_\SM$ and $L_\C$ from the $H\ell\psi\psi\chi$ coupling in~\eqref{eq:Hlqqq} is roughly $T^7 / M^6$. 
Comparing this rate with the expansion rate $\sim T^2 / \Mpl$, one would then get $T_\LD \sim (M / \Mpl)^{1/5} M$.
This is naive, however, as it does not take into account $L_\SM \leftrightarrow L_\C$ shuffling processes due to heavy degrees of freedom responsible for generating \eqref{eq:Hlqqq}.
Those heavy particles can participate in more efficient 2-to-2 scattering processes and may give a much \emph{lower} value of $T_\LD$ than above.
Therefore, $T_\LD$ depends on the unknown UV physics and cannot be estimated from the effective operator~\eqref{eq:Hlqqq}.

As we do not wish to commit to a specific UV model in this work, 
we regard $T_\LD$ as a free parameter hierarchically lower than $M$, i.e., $T_\LD \ll M$ (which is satisfied even by the naive estimate).
We also expect $T_\LD \ll m_\UV$, where $m_\UV$ denotes the mass scale of the unknown UV particles behind the operator~\eqref{eq:Hlqqq}.
This is because LD is triggered when $T$ drops below $m_\UV$ and the number densities of those heavy particles becoming exponentially suppressed.
This gives $T_\LD \sim m_\UV / O(10)$, analogously to the standard estimation of thermal WIMP freeze-out, unless one insists on a really tiny coupling between the SM lepton and heavy particles, defeating the whole point of explaining small neutrino masses by compositeness.

The hierarchy $T_\LD \ll m_\UV$ allows us to determine $c_\LD$ without knowing the unknown UV physics behind the effective coupling~\eqref{eq:Hlqqq}.
Combining~\eqref{eq:leptogenesis} and~\eqref{eq:B-LSM}, we get
\bea
\eta_B^\PS
\simeq
\frac{29}{78} \,c_\LD^\PS \, \frac{T_\LD^2}{f_a^2} \, Y_\theta
\,.\label{eq:etaB}
\eea

Note that PQ charge conservation~\eqref{eq:PQ-charge-relation} is necessary to obtain~\eqref{eq:etaB}.
Since PQ charge conservation requires neglecting the axion potential, we must demand the axion kinetic energy, $f_a^2 \dot\theta^2/2$, to be much larger than the axion potential's height, $\simeq 2f_a^2 m_a^2(T)$, where $m_a(T)$ takes into account the temperature dependence of the potential.
Once the kinetic energy drops below this, the axion field will be trapped. As in~\cite{Co:2019jts}, we define the trapping temperature $T_\trap$ via the relation
\bea
\dot\theta(T_\trap) \equiv 2m_a(T_\trap)
\,,\label{eq:T_trap:def}
\eea
and require $T_\LD \gg T_\trap$. 
As shown in the Supplemental Material, 
this amounts to a much weaker condition than the requirement that LD should occur before the weak sphaleron turns off, i.e., $T_\LD \gg T_\EW$.

\subsection{Dark matter}
In co-genesis, $\dot\theta$ is also responsible for DM production in addition to baryogenesis~\cite{Co:2019wyp,Co:2019jts, Co:2020xlh}. 
This allows the reference to $Y_\theta$ in~\eqref{eq:etaB} to be removed and  
gives rise to a relation between $\eta_B$ and DM abundance. 

\begin{figure}[t] 
\begin{center}
\includegraphics[width=0.45\textwidth]{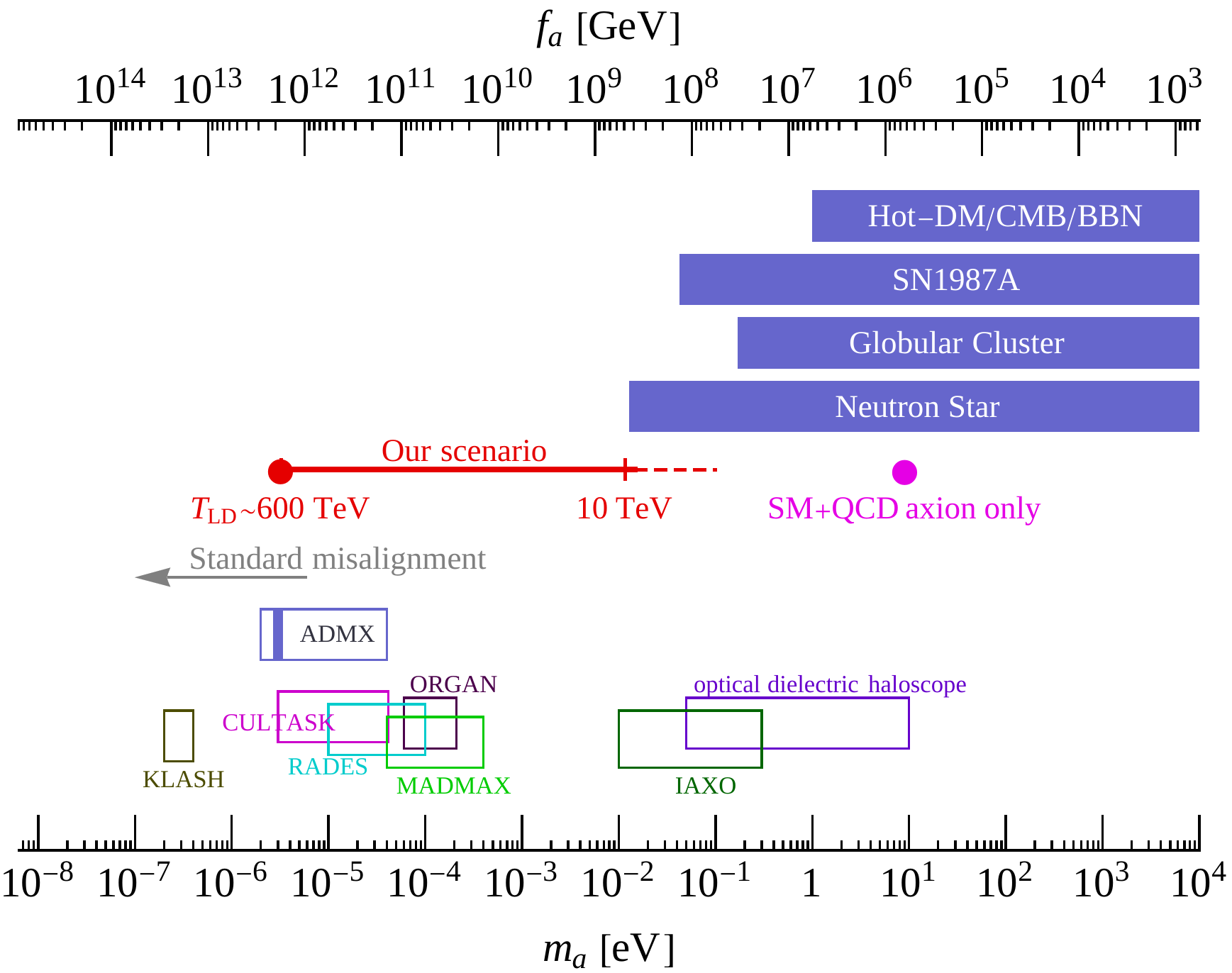} 
\end{center}
\caption{Our prediction for $c_\LD = 0.04$ (red line and dot) is depicted along with current constraints (dark blue shaded bands) and sensitivities of future experiments (open rectangles). The magenta dot represents the minimal co-genesis scenario with just SM + QCD axion. The gray arrow corresponds to the standard misalignment mechanism for DM\@.}
\label{axionplot}
\end{figure}

If $3H(T_\trap) > m_a(T_\trap)$, the axion field will slow-roll until the condition $3H(T) = m_a(T)$ is reached at a temperature $T < T_\trap$ and then will begin to oscillate.
Thus, DM in this case is produced via the standard misalignment mechanism~\cite{Preskill:1982cy,Abbott:1982af,Dine:1982ah,Borsanyi:2016ksw} with an $O(1)$ misalignment angle.
Unlike in the standard case, however, this ``initial'' misalignment angle is not an input parameter in our scenario and cannot be tuned to a small value to avoid a DM overabundance.
Our prediction on $m_a$ and $f_a$ in this case is represented by the red dot in Fig.~\ref{axionplot}.

For $3H(T_\trap) < m_a(T_\trap)$, the axion field immediately begins oscillating around the nearest minimum of the potential, thereby becoming cold DM\@.
This is the ``axion kinetic misalignment'' mechanism of~\cite{Co:2019jts}. 
While the DM number density is given by $sY_\theta$ at $T=T_\trap$, the energy density is not given by $m_a sY_\theta$ because the axion potential is not quite quadratic for an $O(1)$ angle generically expected at the moment of trapping.
A numerical calculation in~\cite{Co:2019jts} shows
\bea
\frac{\rho_\DM^\PS}{s} \simeq 2 m_a^{(0)} Y_\theta
\,,
\label{DM}
\eea
where $m_a^{(0)} \equiv m_a(T)\bigr|_{T=0}$. 
Using this to eliminate $Y_\theta$ in~\eqref{eq:etaB}, we predict that baryon asymmetry and DM abundance should be related as
\bea
\eta_B^\PS
\simeq
\frac{29}{78} \,c_\LD^\PS \, \frac{T_\LD^2}{f_a f_\pi m_\pi} \, \frac{\rho_\DM^\PS}{s}
\,,
\eea
where we have used the leading-order QCD axion relation, $m_a^{(0)} f_a = z m_\pi f_\pi$ with $z \equiv \sqrt{m_u m_d}/(m_u + m_d) \simeq 0.5$.
In terms of $\Omega_B / \Omega_\DM \simeq m_p \eta_B^\PS /(\rho_\DM^\PS/s)$ with the proton mass $m_p$, 
we thus obtain
\bea
\frac{\Omega_B}{\Omega_\DM} 
&\simeq&
\frac{29}{78} \,c_\LD^\PS \, \frac{m_p T_\LD^2}{f_a f_\pi m_\pi}
\nn\\
&\simeq&
0.2 \, \frac{c_\LD^\PS}{0.04} \!\left( \frac{T_\LD}{10\>\TeV} \right)^{\!\!2} \frac{5 \times 10^8\>\GeV}{f_a}
\,.\label{eq:prediction}
\eea
%

As shown in the Supplemental Material, 
 the condition $3H(T_\trap) < m_a(T_\trap)$ translates to $T_\LD \lesssim 300\>\TeV$ for $c_\LD \sim 0.04$ and $g_*^\trap \sim g_{*S}^\trap \sim 100$ (and also for $f_a \sim 10^9\>\GeV$ but with an extremely mild dependence on $f_a$).
With the lower bound $T_\LD > T_\EW \sim 0.1\>\TeV$, we have a wide possible range of $T_\LD$. 

In Fig.~\ref{axionplot}, our prediction~\eqref{eq:prediction} with the benchmark value of $c_\LD = 0.04$ is shown as a red horizontal line segment,
where correct baryon asymmetry and DM abundance are generated, the small Dirac neutrino masses are explained, and the strong CP problem is solved.
As already mentioned, the red dot corresponds to the case where the standard misalignment is responsible for DM rather than kinetic misalignment.
The dark blue shaded bands show constraints from hot DM/CMB/BBN constraints~\cite{Hannestad:2010yi}, SN1987A~\cite{Chang:2018rso, Carenza:2019pxu}, globular clusters~\cite{Ayala:2014pea}, the neutron star cooling rate~\cite{Hamaguchi:2018oqw} (which translates to $T_\LD \gtrsim 10\>\TeV \sqrt{0.04/c_\LD}$), 
and ADMX~\cite{Braine:2019fqb}.
The magenta dot represents the minimal co-genesis model (i.e., just SM + QCD axion) that generates correct baryon asymmetry and DM abundance and solves the strong CP problem, 
but is excluded by the aforementioned bounds except ADMX\@.
The gray arrow describes $m_a$ and $f_a$ for the standard misalignment mechanism for DM with an appropriately tuned initial misalignment angle, where the arrow indicates the direction of more tuning.
It is an exciting prospect that our parameter range will be probed by a variety of future experiments shown by open rectangles in Fig.~\ref{axionplot}~\cite{Braine:2019fqb, McAllister:2017lkb, Melcon:2018dba, Chung:2016ysi,Alesini:2017ifp, TheMADMAXWorkingGroup:2016hpc, Brun:2019lyf, Arvanitaki:2017nhi, Armengaud:2019uso}.

\subsection{$\Delta N_\eff$ and more}
Since we have Dirac neutrinos and our thermal history begins at a temperature above $T_\LD$ where the SM particles and the (constituents of) $\nu^\C$'s are in equilibrium, we have a robust prediction on the lower bound on $\Delta N_\eff$:
\bea
\Delta N_\eff 
\geq \Delta N_\eff \bigr|_\mathrm{min} 
= 3 \!\left( \frac{g_*^\SM(T_{\nu\dd})}{g_*^\SM(T_\kd)} \right)^{\!\! 4/3}
\simeq 0.14
\,,\label{eq:Neff}
\eea
where $T_{\nu\dd} \sim O(1)\>\MeV$ is the neutrino decoupling temperature of the SM, and $T_\kd < T_\LD$ is the temperature at which the SM and composite sectors kinetically decouple from each other.

The absolute lower bound~\eqref{eq:Neff} assumes $T_\kd > m_t$ to include all SM degrees of freedom in $g_*^\SM(T_\kd)$ 
and it is near the edge of the current $1\sigma$ uncertainty from the Planck experiment~\cite{Aghanim:2018eyx}, $N_\eff=2.99 \pm 0.17$.
If $T_\kd$ falls between the charm mass and QCD confinement temperature, 
we would have $\Delta N_\eff \simeq 0.29$, almost at the $2\sigma$ bound.
Thus, near future CMB experiments such as the CMB stage-IV and the CORE mission of the ESA~\cite{Abazajian:2016yjj, Delabrouille:2017rct} are expected to offer firm tests of our scenario.

If $T_\kd > \Lambda_\C$, a large reduction in the degrees of freedom in the composite sector upon confinement would enhance $\Delta N_\eff$ by a factor of $(g_*^{\C+} / g_*^{\C-})^{4/3}$, where $g_*^{\C+(-)}$ is the degrees of freedom in the composite sector above(below) $\Lambda_\C$.
For the $\SU(6)$ benchmark model, 
the enhancement would give  
$\Delta N_\eff \simeq 8.8$, grossly at odds with observation.

Like $T_\LD$, $T_\kd$ depends sensitively on the unknown UV physics behind generating the effective operator~\eqref{eq:Hlqqq} so we treat it as a free parameter.
All we need to show is that $T_\kd$ \emph{can} be lower than $\Lambda_\C$ in some UV models.
Generically, UV dynamics generating~\eqref{eq:Hlqqq} will also generate interactions of the form 
$\bar{\ell} \bar{\sigma}^\mu \ell \, \bar{\Psi} \bar{\sigma}_\mu \Psi / M'^2$ with $\Psi = \psi,\chi$,
as these are allowed by all symmetries of the theory.
For $T \gtrsim \Lambda_\C$, those couplings mediate 2-to-2 processes such as $\ell + \Psi \to \ell + \Psi$ with a rate of order $\sim T^5/(4\pi M'^4)$. 
For $T \lesssim \Lambda_\C$, they mediate $\ell + \nu^\C \to \ell + \nu^\C$, etc., 
which (accidentally) give rates of the same order, $\sim T^5/(4\pi M'^4)$.
Demanding those processes to be in equilibrium, we get $T_\kd \sim (4\pi\sqrt{g_*} M'^4 / \Mpl)^{1/3}$.
Then, by using~\eqref{eq:numass}, the condition $T_\kd \ll \Lambda_\C$ gives
\beq
\Lambda_\C 
\ll \frac{\Mpl}{4\pi\sqrt{g_*}} \!\left( \frac{(4\pi)^2 y_\nu}{N_\C} \right)^{\!\!\frac43} \!\!\left( \frac{M}{M'} \right)^{\!\!4}
\sim 10^2\>\GeV \!\left( \frac{M}{M'} \right)^{\!\!4}
\label{eq:Lcupper}
\eeq
for $g_* \sim 100$ and $y_\nu \simeq 3\times 10^{-13}$.
If $M'=M$, this would imply $T_\kd \ll \Lambda_\C \ll 100\>\GeV$ 
so the absolute lower bound~\eqref{eq:Neff} cannot be reached because $T_\kd$ is below $m_t$,
although there may barely be room for $T_\kd$ to be above the QCD confinement scale to saturate the $2\sigma$ bound. 
However, slight differences in the values of UV couplings can easily give $M$ slightly larger than $M'$ by, e.g., a factor of 5, which will open up a wide gap for $T_\kd$ to sit between $m_t$ and $\Lambda_\C$ and thus lead to $\Delta N_\eff \simeq 0.14$.

However, the analysis above does suggest our scenario will be testable in near future, not only by CMB but also possibly by laboratory experiments. 
If we observe $\Delta N_\eff \simeq 0.14$ or some value close to it in future CMB experiments, 
we should expect a new physics associated with the scale $M'$, followed by a new physics associated with $M$ and the operator~\eqref{eq:Hlqqq}.
For example, in some UV models, operators giving rise to $\mu \to e\gamma$ and/or the electron EDM can be generated along with $\bar{\ell} \bar{\sigma}^\mu \ell \, \bar{\Psi} \bar{\sigma}_\mu \Psi / M'^2$.
The study of such signals is highly UV model dependent by nature, however, and hence beyond the scope of this work.

\section{Summary}
We have shown that the composite origin of small Dirac neutrino masses, when it is combined with a (minimal) QCD axion to solve the strong CP problem, can simultaneously also generate correct baryon asymmetry and DM abundance.
Our parameter space is currently wide open with $m_a$ in the range between $O(10)\>\meV$ and $O(10)\>\mueV$, 
which will be probed in various upcoming experiments. 
Our scenario also predicts a robust lower bound on $\Delta N_\eff$ of 0.14, which, again, will be unambiguously testable in near future. 

\vskip 1em
This work is supported by the US Department of Energy grant DE-SC0010102 and the Japan Society for Promotion of Science (JSPS) grant KAKENHI 21H01086.

\begin{appendix}

\section{Lepton asymmetry}
\label{app:leptogenesis}
Here, we summarize the calculations of lepton asymmetry quoted in the main text.

\subsection{Lagrangian and PQ current}
We start from the basis in which the couplings linear in the axion field $a(x)$ (or $\theta(x) \equiv a(x) / f_a$) take the following form:
\beq
\mathcal{L}_{a\,\text{int}} 
= \frac{\theta}{16\pi^2}  \sum_F c_F^\PS g_F^2 \,\mathrm{tr}[F_{\mu\nu} \widetilde F^{\mu\nu}] 
+ \partial_\mu \theta \sum_f c_f^\PS \bar{f} \bar{\sigma}^\mu f 
\,,\label{eq:axion-int}
\eeq
where $F=1,2,3,6$ respectively indicate the $\mathrm{U}(1)_Y$, $\mathrm{SU}(2)_W$, $\mathrm{SU}(3)_\C$, $\mathrm{SU}(6)$ gauge groups, while $f = q$, $u^\C$, $d^\C$, $\ell$, $e^\C$, $\psi$, $\chi$.
There are no other couplings linear in $\theta$ in this basis;
for example, such coupling to the Higgs doublet has been set to zero without loss of generality by redefining the SM fermion fields by a hypercharge rotation.
We have assumed flavor universality of the axion interactions. 
So, more explicitly, $c_q \bar{q} \sigma^\mu q$ should be interpreted as $\sum_{i=1}^3 c_{q_i} \bar{q}_i \sigma^\mu q_i$ with $c_{q_1} = c_{q_2} = c_{q_3} \equiv c_q$,  $c_\psi \bar{\psi} \sigma^\mu \psi$ as $\sum_{i=1}^2 c_{\psi_i} \bar{\psi}_i \sigma^\mu \psi_i$ with $c_{\psi_1} = c_{\psi_2} \equiv c_\psi$, etc.
Such flavor universality is automatic in both the KSVZ and DFSZ UV constructions unless one chooses to assign flavor-dependent PQ charges.

To simplify our calculations, we will ignore the up-quark Yukawa coupling, $y_u$, in the basis where the quark Yukawa coupling matrices are diagonal.
(Here we strictly mean the 1st-generation up quark so $y_c$ and $y_t$ \emph{will not} be ignored.
For a more precise, but more complicated, formalism with $y_u \neq 0$, see \cite{Domcke:2020kcp}.)
Two things must be mentioned to justify this approximation.
First, the existence of a tiny, nontrivial axion potential is quantitatively irrelevant for the baryogenesis process concerned here as it occurs at much higher energy scales than the height of the potential.
Hence, making the potential flat by setting $y_u = 0$ is a good approximation.
Second, setting $y_u = 0$ does not lead to any additional conservation law.
Each conservation law would act as a constraint upon the minimization of the free energy to be performed below, so we must make sure that our approximation does not introduce an extra constraint that should not exist.
For $y_u$, setting it to zero would perturbatively imply conservation of the $u_1^\C$ number, but the $\mathrm{SU}(3)_\C$ sphaleron violates the $u_1^\C$ number conservation at temperatures relevant for our baryogenesis mechanism.
Therefore, we should be able to capture the physics correctly by \emph{not} requiring the $u_1^\C$ number conservation despite setting $y_u = 0$.

This subtlety matters for the electron, however, because it is possible that the processes that violate the $e^\C_1$ number conservation are not fast enough to be in equilibrium at relevant temperatures.
Below we will analyze two cases depending on whether the $e^\C_1$ number is conserved or not.
  
Now, with $y_u = 0$, we will perform a series of field redefinitions to simplify the axion interactions in~(\ref{eq:axion-int}).
First, we redefine $u^\C_1$ by an appropriate phase to remove the $aG\widetilde{G}$ coupling,
which amounts to the following replacement of coefficients in~(\ref{eq:axion-int}):
\beq
c_3^\PP \to c'_3 = 0
\,,\quad
c_{u^\C_1}^\PP \to c'_{u^\C_1} 
= c_{u^\C_1}^\PP \!- c_3^\PP
= c_{u^\C}^\PP \!- c_3^\PP
\,.\label{eq:c_3-removed}
\eeq
Next, we remove the $aW\widetilde{W}$ coupling by performing the $\mathrm{U}(1)_B$ rotation:
\bea
c_2^\PP &\to& c'_2 = 0 \,, \\   
c_q^\PP &\to& c'_q = c_q^\PP - \frac{c_2}{9} \,, \\ 
c'_{u^\C_1} &\to& c''_{u^\C_1} 
= c'_{u^\C_1} + \frac{c_2}{9}
= c_{u^\C}^\PP \!- c_3^\PP + \frac{c_2}{9}
\,, \\ 
c_{u^\C_{2,3}}^\PP &\to& c'_{u^\C_{2,3}} 
= c_{u^\C}^\PP \!+ \frac{c_2}{9} 
\,,\\ 
c_{d^\C}^\PP &\to& c'_{d^\C} 
= c_{d^\C}^\PP \!+ \frac{c_2}{9}
\,.\label{eq:c_2-removed}
\eea
Finally, to rotate away the $aG_6\widetilde{G}_6$ term, 
we need an accidental global $\mathrm{U}(1)$ symmetry that is anomalous under the $\mathrm{SU}(6)$ gauge interaction.
Inspecting all the interactions of our entire SM+composite Lagrangian described in the main text, 
we see that such $\mathrm{U}(1)$ symmetry indeed exists, 
where $\psi$ and $\chi$ carry charges $1$ and $-2$, respectively, and no SM fields nor axion are charged. 
Being anomalous under $\mathrm{SU}(6)$, a rotation in this $\mathrm{U}(1)$ can be used to remove the $aG_6\widetilde{G}_6$ term as
\beq
c_6^\PP \to c'_6 = 0
\,,\quad 
c_{\psi}^\PP \to c'_{\psi} = c_\psi^\PP \!+ \frac{c_6}{6}
\,,\quad 
c_{\chi}^\PP \to c'_{\chi} = c_\chi^\PP \!- \frac{c_6}{3}
\,.\label{eq:c_6-removed}
\eeq
We do not have an anomalous global symmetry we could use to remove the $aB\widetilde{B}$ term.
Nevertheless, since $\int\!\dd^4x\, B\widetilde{B}$ is a vanishing surface integral, 
the presence of $aB\widetilde{B}$ will have no effects on our baryogenesis mechanism.
If the $e^\C_1$ number violating interactions are too slow to be in equilibrium, 
we can set $y_e = 0$ and this will in turn allow us to rotate away $aB\widetilde{B}$ by redefining $e^\C_1$.
This might appear to contradict the statement that $B\widetilde{B}$ is irrelevant, because $c_1$ will still be in the Lagrangian as part of $c'_{e^\C_1}$, just as $c_3$ is part of $c'_{u^\C_1}$.
But precisely because of the $e^\C_1$ conservation that needs to be imposed in this case, $c'_{e^\C_1}$ will drop out from our calculation and hence the contradiction will not arise.

Our Lagrangian is now manifestly invariant under a global shift of the axion field, $\theta(x) \to \theta(x) + \alpha$ with a spacetime constant $\alpha$, except for the $aB\widetilde{B}$ term.
This implies the conservation law:
\bea
\partial_\mu J^\mu_\PQ = c'_1 \frac{g_1^2}{16\pi^2} B_{\mu\nu} \widetilde{B}^{\mu\nu}
\label{eq:PQ-conservation:local}
\eea
with
\beq
J^\mu_\PQ 
= f_a^2 \,\partial^{\mu} \theta
+\sum_f c_f^\PS \bar{f} \bar{\sigma}^\mu f 
\quad\text{with replacements~(\ref{eq:c_3-removed})--(\ref{eq:c_6-removed})} 
\,.\label{eq:PQ-current}
\eeq
We do not need an expression for $c'_1$ as the right-hand side of~(\ref{eq:PQ-conservation:local}) will vanish when integrated over spacetime, implying that the total PQ charge, $\int\!\dd^3x \, J^0_\PQ$, is conserved regardless of the value of $c'_1$. 
If we had chosen not to neglect $y_u$, a term of the form $\I y_u H q u^\C + \mathrm{c.c.}$ would appear on the right-hand side above and $J^\mu_\PQ$ would no longer be conserved.

We assume that the fermions' and scalars' number densities as well as the axion field are spatially homogeneous throughout the baryogenesis process, so we always have $J^{1,2,3}_\PQ = 0$. 
We also assume vanishing initial number densities for all the $f$'s and $\Phi$'s, 
so the initial condition is given by $J^0_\PQ = f_a^2 \dot\theta_\I$.
Then, the PQ charge conservation implies that at later times we have
\beq
\hspace{-0.2cm}
R_\I^3 \, f_a^2 \dot\theta_\I
= R^3 \Bigl( f_a^2 \dot\theta
+\sum_f c_f^\PS n_f^\PS
\Bigr)
\quad\text{with replacements~(\ref{eq:c_3-removed})--(\ref{eq:c_6-removed})}
\,,  
\eeq
where $R$ ($R_\I$) is the (initial) scale factor of the expanding universe, 
and $n_f \equiv \bar{f} \bar{\sigma}^0 f$.
Assuming entropy conservation $s_\I R_\I^3 = s R^3$, this becomes
\bea
\frac{f_a^2 \dot\theta_\I}{s_\I}
= \frac{f_a^2 \dot\theta + n_\PQ^\PP}{s}
\,,\label{eq:PQ-conservation:global}
\eea
with
\bea
n_\PQ^\PP
&\equiv& 
 \Bigl( c_{u^\C} + \frac{c_2}{9} - c_3 \Bigr) n_{u^\C_1}^\PP
+\Bigl( c_{u^\C} + \frac{c_2}{9} \Bigr) n_{u^\C_{23}}^\PP \nn \\
&&
+\Bigl( c_{d^\C} + \frac{c_2}{9} \Bigr) n_{d^\C}^\PP 
+ \Bigl( c_q - \frac{c_2}{9} \Bigr) n_q \nn \\
&&
+c_\ell^\PP n_{\ell}^\PP + c_{e^\C} n_{e^\C}
\nn\\
&&
+\Bigl( c_\psi + \frac{c_6}{6} \Bigr) n_\psi
+\Bigl( c_\chi - \frac{c_6}{3} \Bigr) n_\chi
\,,\label{eq:n_PQ}
\eea
where 
\bea
n_f^\PP &\equiv& n_{f_1}^\PP \!+ n_{f_2}^\PP \!+ n_{f_3}^\PP 
\>\>\text{for $f=q$, $d^\C$, $\ell$, $e^\C$}  \,,\nn\\
n_{u^\C_{23}} &\equiv& n_{u^\C_2} \!+ n_{u^\C_3}  \,,\quad
n_\psi \equiv n_{\psi_1} \!+ n_{\psi_2}  \,.
\eea

\subsection{Minimizing the free energy and determining the asymmetries}
We will determine the values of the number densities in~(\ref{eq:PQ-conservation:global}) by minimizing the free energy.
We write the free energy density $\mathcal{F}$ as
\bea
\mathcal{F} 
= \rho_{a} + \rho_\text{p} - T s_\text{p}
\,,\label{eq:Free-Energy:def}
\eea
where $\rho_a = f_a^2 \dot\theta^2 / 2$ is the energy density of the classical motion of the axion field, while $\rho_\text{p}$ and $s_\text{p}$ are respectively the energy and entropy densities of the plasma.
Using~(\ref{eq:PQ-conservation:global}) and~(\ref{eq:n_PQ}), we can express $\rho_a$ in terms of the number densities as
\bea
\rho_a = \frac{1}{2f_a^2} \!\left( \frac{s}{s_\I} f_a^2 \dot\theta_\I - n_\PQ^\PP \right)^{\!\!2}.
\label{eq:rho_a} 
\eea
We will also express $\rho_\text{p}$ and $s_\text{p}$ in terms of the number densities below.
Then, $\mathcal{F}$ will be a function of the number densities, and the values of the number densities at equilibrium should minimize $\mathcal{F}$. 
The minimization determines the number densities completely in terms of the $c$ coefficients and $\dot\theta$ without needing to solve a large number of Boltzmann equations as in Refs.~\cite{Co:2019wyp, Domcke:2020kcp}.
This simplicity is made possible because of the PQ charge conservation~(\ref{eq:PQ-conservation:global}) coming from our approximation of setting $y_u = 0$.
For $y_u \neq 0$, the Boltzmann equations would be needed. 

To calculate $\rho_\text{p}$ and $s_\text{p}$ in terms of the number densities,
let us first work out relations between the number density, $n_i$, and chemical potential, $\mu_i$, of a particle species $i$.
Neglecting interactions between particles, for a free massless fermion $f$ with degeneracy $N_f$ in equilibrium at $T \gg \mu_f$, we have
\bea
n_{f} &=& N_f \!\int\!\! \frac{\dd^3 p}{(2\pi)^3} \frac{1}{e^{(E-\mu_f)/T}+1} - (\mu_f \to -\mu_f) \nn \\
&=& \frac{N_f}{6}\mu_f T^2 + O(\mu_f^3) \,.
\label{nf}
\eea
Similarly, for a free massless scalar boson $b$ with degeneracy $N_b$, we have
\bea
n_{b} &=& N_b \!\int\!\! \frac{\dd^3 p}{(2\pi)^3} \frac{1}{e^{(E-\mu_b)/T}-1} - (\mu_b \to -\mu_b) \nn\\
&=& \frac{N_b}{3}\mu_b T^2 + O(\mu_b^3) \,. 
\label{nb}
\eea
Therefore, neglecting the $O(\mu_{f,b}^3)$ corrections, we have
\bea
\mu_f  = \frac{6n_{f}}{N_f T^2} \,,\quad
\mu_b = \frac{3 n_{b}}{N_b T^2} \,.
\eea
Using these, we can express the energy densities of $f$ and $b$ in terms of $n_f$ and $n_b$ as
\bea
\rho_f 
&=& N_f \!\int\!\! \frac{\dd^3 p}{(2\pi)^3} \frac{E}{e^{(E-\mu_f)/T}+1} + (\mu_f \to -\mu_f) \nn \\
&=& N_f \!\left( \frac{7}{8}  \frac{\pi^2}{15} T^4 + \frac{1}{4}\mu_f^2 T^2 \right)\! + O(\mu_f^4) \nn\\
&=& \frac{\pi^2}{15} \frac{7N_f }{8} T^4 + 9 \frac{n_{f}^2}{N_f T^2} + O(n_f^4)
\,,
\eea
\bea
\rho_b 
&=& N_b \!\int\!\! \frac{\dd^3 p}{(2\pi)^3} \frac{E}{e^{(E-\mu_f)/T}-1} + (\mu_b \to -\mu_b) \nn\\
&=& N_b \!\left( \frac{\pi^2}{15} T^4 + \frac{1}{2}\mu_f^2 T^2\right)\! + O(\mu_b^4) \nn\\
&=& \frac{\pi^2}{15} N_b T^4 + \frac{9}{2}\frac{n_{b}^2}{N_b T^2} + O(n_b^4) \,.
\eea
Thus, up to $O(n_{f,b}^4)$ corrections, the total energy density of the plasma becomes
\bea
\rho_\text{p} = \frac{\pi^2}{30}g_* T^4+  9\sum_f \frac{ n_{f}^2}{N_f T^2}+ \frac{9}{2} \sum_b \frac{ n_{b}^2}{N_b T^2}
\eea
where
\bea
f &=& q, u^\C_1, u^\C_{23}, d^\C, \ell, e^\C, \psi, \chi   \nn\\
&&\text{with}\quad
(N_q, N_{u^\C_1}, N_{u^\C_{23}}, N_{d^\C}, N_\ell, N_{e^\C}, N_\psi, N_\chi) \nn\\
&& ~~~~~~~~~~~~~~~~~~~~~~~~~~~=  (18,3,6,9,6,3,12,15)
\,,\nn\\
b &=& H \quad\text{with}\quad
N_H = 2
\,.\label{eq:fb}
\eea
(If the $e^\C_1$ number violating processes are out of equilibrium, 
we would need to separate out $e^\C_1$ from $e^\C_{2,3}$ set $n_{e^\C_1} = 0$, 
but this can be effectively achieved without altering the notations by simply changing $N_{e^\C}$ from 3 to 2.)    
Therefore, from $\rho - Ts + P = \sum n\mu$ with $P=\rho/3$ for free massless particles, we have
\bea
\rho_\text{p} - T s_\text{p} &=&  -\frac{\rho_\text{p}}{3} + \sum_{i=f,b} n_i \mu_i  \\
&=& -\frac{\pi^2}{90} g_* T^4 + 3\sum_f \frac{n_{f}^2}{N_f T^2} + \frac{3}{2} \sum_b \frac{n_{b}^2}{N_b T^2}
\,. \nn
\eea
Combining this with~(\ref{eq:rho_a}) as in (\ref{eq:Free-Energy:def}), we finally obtain
\beq
\mathcal{F} = 
\frac{1}{2f_a^2} \!\left( \frac{s}{s_\I}  f_a^2 \dot\theta_\I - n_\PQ^\PP \right)^{\!\!2}
+ 3\sum_f \frac{n_{f}^2}{N_f T^2} + \frac{3}{2} \sum_b \frac{n_{b}^2}{N_b T^2} -\frac{\pi^2}{90} g_* T^4
\,,\label{eq:Free-Energy}
\eeq
where $f,b$ are given in~(\ref{eq:fb}).

Now, the minimization of $\mathcal{F}$ in (\ref{eq:Free-Energy}) is subject to constraints due to conservation laws. 
For example, the total hypercharge must be zero, $\langle Y \rangle = 0$, at all temperatures.
There can be other conserved quantities depending on $T$.
The conservation/violation of the $e^\C_1$ number must be also taken into account.
We consider each case separately below:

\underline{\textbf{(i) $T_\LD \lesssim T \ll m_\UV $, $e^\C_1$ number not conserved:}}
In this case, the only conserved quantities are $Y$ and $B-L$ as the processes converting between $L_\SM$ and $L_\C$ are still in equilibrium. 
So, we minimize $\mathcal{F}$  with respect to the number densities subject to these two constraints:
\begin{widetext}
\bea
0 &=& \langle Y \rangle 
= \frac16 n_q - \frac23 (n_{u^\C_1} + n_{u^\C_{23}}) + \frac13 n_{d^\C} - \frac12 n_\ell + n_{e^\C} + \frac12 n_H  
\,,\nn\\
0 &=& \langle B-L \rangle 
= \frac13 (n_q - n_{u^\C_1} - n_{u^\C_{23}} - n_{d^\C}) - (n_\ell - n_{e^\C}) 
-\Bigl( -\frac23 n_{\psi} + \frac13 n_{\chi}  \Bigr)
\,.
\eea
For $T \ll f_a$, the results are:
\bea
\frac{\langle B \rangle}{s} = \frac{\langle L \rangle}{s}
= \Bigl( 
\frac{14}{117} c_3 - \frac{62}{351} c_2 - \frac{7}{78} c_6
+ \frac{31}{39} c_q - \frac{14}{39} c_{u^\C} - \frac{17}{39} c_{d^\C} 
&+& \frac{8}{39} c_\ell - \frac{3}{26} c_{e^\C}
\nn\\
&-& \frac{28}{117} c_{\psi} + \frac{35}{234} c_{\chi} 
\Bigr) \frac{\dot\theta_\I}{s_\I} T^2 
\,,
\eea
and
\bea
\frac{\langle B-L_\SM \rangle}{s} = \frac{\langle L_\C \rangle}{s}
= \Bigl( 
-\frac{35}{936} c_3 - \frac{49}{702} c_2 - \frac{79}{312} c_6 
+ \frac{49}{156} c_q + \frac{35}{312} c_{u^\C} - \frac{133}{312} c_{d^\C} 
&-& \frac{49}{156} c_\ell + \frac{7}{104} c_{e^\C}
\nn\\
&-& \frac{79}{117} c_{\psi} + \frac{395}{936} c_{\chi} 
\Bigr) \frac{\dot\theta_\I}{s_\I} T^2 
\,.
\eea

\underline{\textbf{(i') $T_\LD \lesssim T \ll m_\UV $, $e^\C_1$ number conserved:}}
As explained earlier,
this case can be obtained by redoing case~(i) with $N_{e^\C} = 2$ instead of 3.
The results for $T \ll f_a$ are:
\bea
\frac{\langle B \rangle}{s} = \frac{\langle L \rangle}{s} 
= \Bigl( 
\frac{95}{846} c_3 - \frac{220}{1269} c_2 - \frac{13}{141} c_6
+ \frac{110}{141} c_q - \frac{95}{282} c_{u^\C} - \frac{125}{282} c_{d^\C} 
&+& \frac{31}{141} c_\ell - \frac{4}{47} c_{e^\C}
\nn\\
&-& \frac{104}{423} c_{\psi} + \frac{65}{423} c_{\chi} 
\Bigr) \frac{\dot\theta_\I}{s_\I} T^2 
\,,
\eea
and
\bea
\frac{\langle B-L_\SM \rangle}{s} = \frac{\langle L_\C \rangle}{s}
= \Bigl( 
-\frac{14}{423} c_3 - \frac{91}{1269} c_2 - \frac{71}{282} c_6 
+ \frac{91}{282} c_q + \frac{14}{141} c_{u^\C} - \frac{119}{282} c_{d^\C} 
&-& \frac{91}{282} c_\ell + \frac{7}{141} c_{e^\C}
\nn\\
&-& \frac{284}{423} c_{\psi} + \frac{355}{846} c_{\chi} 
\Bigr) \frac{\dot\theta_\I}{s_\I} T^2 
\,.
\eea

\underline{\textbf{(ii) $T \ll T_\LD$, $e^\C_1$ number not conserved:}}
At these temperatures, the processes converting between $L_\SM$ and $L_\C$ have frozen out.
Therefore, we now have three constraints to be imposed upon minimizing $\mathcal{F}$:
\bea
\langle Y \rangle = 0
\,,\quad
\frac{\langle B-L_\SM \rangle}{s} = \frac{\langle B-L_\SM \rangle}{s} \biggr|_\LD
\,,\quad
\frac{\langle L_\C \rangle}{s} = \frac{\langle B-L_\SM \rangle}{s} \biggr|_\LD
\,,
\eea
where $\langle B-L_\SM \rangle_\LD^\PD$ is the frozen-out value of $\langle B-L_\SM \rangle$ at the lepton number decoupling.
Minimizing $\mathcal{F}$ under these constraints, we obtain
\bea
\frac{\langle B \rangle}{s}
= \frac{28}{79} \frac{\langle B-L_\SM \rangle}{s} \biggr|_\LD
+\Bigl( 
\frac{21}{158} c_3 - \frac{12}{79} c_2
+ \frac{54}{79} c_q - \frac{63}{158} c_{u^\C} - \frac{45}{158} c_{d^\C} 
+ \frac{25}{79} c_\ell - \frac{11}{79} c_{e^\C} 
\Bigr) \frac{\dot\theta_\I}{s_\I} T^2 
\,.
\eea
\end{widetext}
Evaluating this expression at $T = T_\EW$ at which the $\mathrm{SU}(2)_W$ sphalerons freeze out provides us with the final value of $\langle B \rangle$.
We ignore the second term above as its $T^2$ scaling implies that it is negligible compared to the first term.

\underline{\textbf{(ii') $T \ll T_\LD$, $e^\C_1$ number conserved:}}
There is no need to analyze this case, because at temperatures near, but still above, the electroweak scale, the $e^\C_1$ number violating interactions mediated by $y_e$ are in equilibrium.

\section{Conservation of $Y_\theta$}
\label{app:Ythetaconservation}
The results above tell us that
\bea
\frac{n_\PQ^\PP}{s} \sim T^2 \frac{\dot\theta_\I}{s_\I}
\ll f_a^2 \frac{\dot\theta_\I}{s_\I}
\eea
for $T \ll f_a$.
Then, since we are always in this temperature regime, the $n_\PQ$ term in the PQ charge conservation relation~\eqref{eq:PQ-conservation:global} is negligible and hence $Y_\theta$ is conserved to an excellent approximation.
This is valid under the condition $T \gg T_\trap$ so that the axion potential can be ignored to justify PQ charge conservation~\eqref{eq:PQ-conservation:global} in the first place.

\section{Trapping temperature}
\label{app:trapping}
Here, we estimate $T_\trap$ defined in~\eqref{eq:T_trap:def} and analyze what the conditions $T_\LD \gg T_\trap$ and $3H(T_\trap) \ll m_a(T_\trap)$ imply.

First, we want to relate $\dot\theta(T_\trap)$ in~\eqref{eq:T_trap:def} to the observed baryon asymmetry, $\eta_B \simeq 8.5 \times 10^{-11}$.
Using the conservation of $Y_\theta$, which we expect to be still roughly valid at $T_\trap$,
we obtain $\dot\theta(T_\trap)$ from the relation~\eqref{eq:etaB}:
\bea
\dot\theta (T_\trap) \simeq \frac{78}{29} \frac{2\pi^2}{45} \frac{\eta_B^\PS \, g_{*S}^\trap}{c_\LD} \frac{T_\trap^3}{T_\LD^2}
\label{thetadotTld} 
\eea
Next, we want to estimate $m_a(T_\trap)$ in~\eqref{eq:T_trap:def}.
Since our axion is solving the strong CP problem, 
the axion potential $\sim f_a^2 m_a^2(T)$ is generated from QCD\@.
A recent lattice calculation~\cite{Borsanyi:2016ksw} has shown that the potential scales as $(T_\QCD/T)^{2n}$ for $T>T_\QCD$ with $T_\QCD \simeq 150\>\MeV$ and $n \simeq 4$.
Combining all these, \eqref{eq:T_trap:def} gives
\bea
T_\trap 
&\sim&
\!\left( 
\frac{2\cdot29}{78} \frac{45}{2\pi^2} \frac{c_\LD}{\eta_B^\PS \, g_{*S}^\trap} m_a^{(0)} T_\QCD^n T_\LD^2 
\right)^{\!\!\frac{1}{n+3}} \nn\\
&\sim&
\!\left( 
\frac{c_\LD}{\eta_B^\PS \, g_{*S}^\trap} \frac{T_\QCD^{n+2} T_\LD^2}{f_a} 
\right)^{\!\!\frac{1}{n+3}},
\label{eq:T_trap}
\eea
where we have used $m_a^{(0)} \simeq z m_\pi f_\pi / f_a \simeq 0.25 \, T_\QCD^2/f_a$ and dropped the factor of $(0.25\frac{2\cdot29}{78}\frac{45}{2\pi^2})^{1/(n+3)} \simeq 1$ in the last step.
With this, the condition $T_\trap \gg T_\QCD$ to justify the use of the scaling $m_a(T) \propto T^{-n}$ is equivalent to
\beq
T_\LD
\gg
\sqrt{ \frac{\eta_B^\PS \, g_{*S}^\trap}{c_\LD} T_\QCD f_a }
\sim
10\>\GeV \, \sqrt{\frac{g_{*S}^\trap}{100} \frac{0.04}{c_\LD} \frac{f_a}{10^9\>\GeV} }
\,.
\eeq
This condition is always satisfied because it is weaker than the condition that lepton-number decoupling should occur before the weak sphaleron turns off, i.e., $T_\LD \gg T_\EW \sim 100\>\GeV$.

To justify our use of PQ charge conservation during leptogenesis, 
we must require $T_\LD \gg T_\trap$. 
With~\eqref{eq:T_trap}, this is equivalent to imposing that
\bea
T_\LD
&\gg&
\!\left(
\frac{c_\LD}{\eta_B^\PS \, g_{*S}^\trap} \frac{T_\QCD}{f_a} 
\right)^{\!\!\frac{1}{n+1}} \! T_\QCD \nn\\
&\sim&
40\>\MeV
\left(
\frac{c_\LD}{0.04} \frac{100}{g_{*S}^\trap} \frac{10^9\>\GeV}{f_a} 
\right)^{\!\!\frac{1}{5}} ,
\eea
which, again, is already satisfied.

Next, we need to require $3H(T_\trap) \ll m_a(T_\trap)$ as explained the main text.
For a radiation dominated universe, this condition means 
\bea
\frac{\sqrt{g_*^\trap} T_\trap^2}{\Mpl} \ll m_a^{(0)} \!\left( \frac{T_\QCD}{T_\trap} \right)^{\!\! n}
\,,
\eea
where $\Mpl\simeq 2.4\times 10^{18}\,\GeV$.
Using the middle expression of~\eqref{eq:T_trap} with $m_a^{(0)} \simeq 0.25 \, T_\QCD^2/f_a$ to eliminate $T_\trap$, this amounts to the following upper bound on $T_\LD$: 
\bea
T_\LD
&\ll& 
6.8
\!\left( \frac{\eta_B^\PS \, g_{*S}^\trap}{c_\LD} T_\QCD \right)^{\!\!\frac{n+2}{2n+4}} 
\!\left( \frac{\Mpl}{\sqrt{g_*^\trap}} \right)^{\!\!\frac{n+3}{2n+4}} 
\!\left( \frac{1}{f_a} \right)^{\!\!\frac{1}{2n+4}}
\nn\\
&\sim&
3 \times 10^2\>\TeV 
\left( \frac{0.04}{c_\LD} \frac{g_{*S}^\trap}{100} \right)^{\!\!\frac12}
\!\left( \frac{100}{g_*^\trap} \right)^{\!\!\frac{7}{24}}
\!\left( \frac{10^9\>\GeV}{f_a} \right)^{\!\!\frac{1}{12}} . \nn\\
\eea 

\end{appendix}


\begin{thebibliography}{10}

\bibitem{ArkaniHamed:1998pf}
N.~Arkani-Hamed and Y.~Grossman, ``{Light active and sterile neutrinos from
  compositeness},'' \href{http://dx.doi.org/10.1016/S0370-2693(99)00672-3}{{\em
  Phys. Lett. B} {\bfseries 459} (1999) 179--182},
  \href{http://arxiv.org/abs/hep-ph/9806223}{{\ttfamily arXiv:hep-ph/9806223}}.

\bibitem{Gherghetta:2003hf}
T.~Gherghetta, ``{Dirac neutrino masses with Planck scale lepton number
  violation},'' \href{http://dx.doi.org/10.1103/PhysRevLett.92.161601}{{\em
  Phys. Rev. Lett.} {\bfseries 92} (2004) 161601},
  \href{http://arxiv.org/abs/hep-ph/0312392}{{\ttfamily arXiv:hep-ph/0312392}}.

\bibitem{Co:2019wyp}
R.~T. Co and K.~Harigaya, ``{Axiogenesis},''
  \href{http://dx.doi.org/10.1103/PhysRevLett.124.111602}{{\em Phys. Rev.
  Lett.} {\bfseries 124} no.~11, (2020) 111602},
  \href{http://arxiv.org/abs/1910.02080}{{\ttfamily arXiv:1910.02080
  [hep-ph]}}.

\bibitem{Co:2019jts}
R.~T. Co, L.~J. Hall, and K.~Harigaya, ``{Axion Kinetic Misalignment
  Mechanism},'' \href{http://dx.doi.org/10.1103/PhysRevLett.124.251802}{{\em
  Phys. Rev. Lett.} {\bfseries 124} no.~25, (2020) 251802},
  \href{http://arxiv.org/abs/1910.14152}{{\ttfamily arXiv:1910.14152
  [hep-ph]}}.

\bibitem{Co:2020xlh}
R.~T. Co, L.~J. Hall, and K.~Harigaya, ``{Predictions for Axion Couplings from
  ALP Cogenesis},'' \href{http://dx.doi.org/10.1007/JHEP01(2021)172}{{\em JHEP}
  {\bfseries 01} (2021) 172}, \href{http://arxiv.org/abs/2006.04809}{{\ttfamily
  arXiv:2006.04809 [hep-ph]}}.

\bibitem{Dimopoulos:1980hn}
S.~Dimopoulos, S.~Raby, and L.~Susskind, ``{Light Composite Fermions},''
  \href{http://dx.doi.org/10.1016/0550-3213(80)90215-1}{{\em Nucl. Phys. B}
  {\bfseries 173} (1980) 208--228}.

\bibitem{Csaki:2021xhi}
C.~Cs\'aki, H.~Murayama, and O.~Telem, ``{Some Exact Results in Chiral Gauge
  Theories},'' \href{http://arxiv.org/abs/2104.10171}{{\ttfamily
  arXiv:2104.10171 [hep-th]}}.

\bibitem{Smith:2021vbf}
P.~B. Smith, A.~Karasik, N.~Lohitsiri, and D.~Tong, ``{On Discrete Anomalies in
  Chiral Gauge Theories},'' \href{http://arxiv.org/abs/2106.06402}{{\ttfamily
  arXiv:2106.06402 [hep-th]}}.

\bibitem{DOnofrio:2014rug}
M.~D'Onofrio, K.~Rummukainen, and A.~Tranberg, ``{Sphaleron Rate in the Minimal
  Standard Model},''
  \href{http://dx.doi.org/10.1103/PhysRevLett.113.141602}{{\em Phys. Rev.
  Lett.} {\bfseries 113} no.~14, (2014) 141602},
  \href{http://arxiv.org/abs/1404.3565}{{\ttfamily arXiv:1404.3565 [hep-ph]}}.

\bibitem{Co:2020jtv}
R.~T. Co, N.~Fernandez, A.~Ghalsasi, L.~J. Hall, and K.~Harigaya,
  ``{Lepto-Axiogenesis},''
  \href{http://dx.doi.org/10.1007/JHEP03(2021)017}{{\em JHEP} {\bfseries 21}
  (2020) 017}, \href{http://arxiv.org/abs/2006.05687}{{\ttfamily
  arXiv:2006.05687 [hep-ph]}}.

\bibitem{Harigaya:2021txz}
K.~Harigaya and R.~Wang, ``{Axiogenesis from $SU(2)_R$ phase transition},''
  \href{http://arxiv.org/abs/2107.09679}{{\ttfamily arXiv:2107.09679
  [hep-ph]}}.

\bibitem{Preskill:1982cy}
J.~Preskill, M.~B. Wise, and F.~Wilczek, ``{Cosmology of the Invisible
  Axion},'' \href{http://dx.doi.org/10.1016/0370-2693(83)90637-8}{{\em Phys.
  Lett. B} {\bfseries 120} (1983) 127--132}.

\bibitem{Abbott:1982af}
L.~F. Abbott and P.~Sikivie, ``{A Cosmological Bound on the Invisible Axion},''
  \href{http://dx.doi.org/10.1016/0370-2693(83)90638-X}{{\em Phys. Lett. B}
  {\bfseries 120} (1983) 133--136}.

\bibitem{Dine:1982ah}
M.~Dine and W.~Fischler, ``{The Not So Harmless Axion},''
  \href{http://dx.doi.org/10.1016/0370-2693(83)90639-1}{{\em Phys. Lett. B}
  {\bfseries 120} (1983) 137--141}.

\bibitem{Borsanyi:2016ksw}
S.~Borsanyi {\em et~al.}, ``{Calculation of the axion mass based on
  high-temperature lattice quantum chromodynamics},''
  \href{http://dx.doi.org/10.1038/nature20115}{{\em Nature} {\bfseries 539}
  no.~7627, (2016) 69--71}, \href{http://arxiv.org/abs/1606.07494}{{\ttfamily
  arXiv:1606.07494 [hep-lat]}}.

\bibitem{Hannestad:2010yi}
S.~Hannestad, A.~Mirizzi, G.~G. Raffelt, and Y.~Y.~Y. Wong, ``{Neutrino and
  axion hot dark matter bounds after WMAP-7},''
  \href{http://dx.doi.org/10.1088/1475-7516/2010/08/001}{{\em JCAP} {\bfseries
  08} (2010) 001}, \href{http://arxiv.org/abs/1004.0695}{{\ttfamily
  arXiv:1004.0695 [astro-ph.CO]}}.

\bibitem{Chang:2018rso}
J.~H. Chang, R.~Essig, and S.~D. McDermott, ``{Supernova 1987A Constraints on
  Sub-GeV Dark Sectors, Millicharged Particles, the QCD Axion, and an
  Axion-like Particle},'' \href{http://dx.doi.org/10.1007/JHEP09(2018)051}{{\em
  JHEP} {\bfseries 09} (2018) 051},
  \href{http://arxiv.org/abs/1803.00993}{{\ttfamily arXiv:1803.00993
  [hep-ph]}}.

\bibitem{Carenza:2019pxu}
P.~Carenza, T.~Fischer, M.~Giannotti, G.~Guo, G.~Mart\'\i{}nez-Pinedo, and
  A.~Mirizzi, ``{Improved axion emissivity from a supernova via nucleon-nucleon
  bremsstrahlung},''
  \href{http://dx.doi.org/10.1088/1475-7516/2019/10/016}{{\em JCAP} {\bfseries
  10} no.~10, (2019) 016}, \href{http://arxiv.org/abs/1906.11844}{{\ttfamily
  arXiv:1906.11844 [hep-ph]}}. [Erratum: JCAP 05, E01 (2020)].

\bibitem{Ayala:2014pea}
A.~Ayala, I.~Dom\'\i{}nguez, M.~Giannotti, A.~Mirizzi, and O.~Straniero,
  ``{Revisiting the bound on axion-photon coupling from Globular Clusters},''
  \href{http://dx.doi.org/10.1103/PhysRevLett.113.191302}{{\em Phys. Rev.
  Lett.} {\bfseries 113} no.~19, (2014) 191302},
  \href{http://arxiv.org/abs/1406.6053}{{\ttfamily arXiv:1406.6053
  [astro-ph.SR]}}.

\bibitem{Hamaguchi:2018oqw}
K.~Hamaguchi, N.~Nagata, K.~Yanagi, and J.~Zheng, ``{Limit on the Axion Decay
  Constant from the Cooling Neutron Star in Cassiopeia A},''
  \href{http://dx.doi.org/10.1103/PhysRevD.98.103015}{{\em Phys. Rev. D}
  {\bfseries 98} no.~10, (2018) 103015},
  \href{http://arxiv.org/abs/1806.07151}{{\ttfamily arXiv:1806.07151
  [hep-ph]}}.

\bibitem{Braine:2019fqb}
{\bfseries ADMX} Collaboration, T.~Braine {\em et~al.}, ``{Extended Search for
  the Invisible Axion with the Axion Dark Matter Experiment},''
  \href{http://dx.doi.org/10.1103/PhysRevLett.124.101303}{{\em Phys. Rev.
  Lett.} {\bfseries 124} no.~10, (2020) 101303},
  \href{http://arxiv.org/abs/1910.08638}{{\ttfamily arXiv:1910.08638
  [hep-ex]}}.

\bibitem{McAllister:2017lkb}
B.~T. McAllister, G.~Flower, E.~N. Ivanov, M.~Goryachev, J.~Bourhill, and M.~E.
  Tobar, ``{The ORGAN Experiment: An axion haloscope above 15 GHz},''
  \href{http://dx.doi.org/10.1016/j.dark.2017.09.010}{{\em Phys. Dark Univ.}
  {\bfseries 18} (2017) 67--72},
  \href{http://arxiv.org/abs/1706.00209}{{\ttfamily arXiv:1706.00209
  [physics.ins-det]}}.

\bibitem{Melcon:2018dba}
A.~A. Melc\'on {\em et~al.}, ``{Axion Searches with Microwave Filters: the
  RADES project},'' \href{http://dx.doi.org/10.1088/1475-7516/2018/05/040}{{\em
  JCAP} {\bfseries 05} (2018) 040},
  \href{http://arxiv.org/abs/1803.01243}{{\ttfamily arXiv:1803.01243
  [hep-ex]}}.

\bibitem{Chung:2016ysi}
W.~Chung, ``{CULTASK, The Coldest Axion Experiment at CAPP/IBS in Korea},''
  \href{http://dx.doi.org/10.22323/1.263.0047}{{\em PoS} {\bfseries CORFU2015}
  (2016) 047}.

\bibitem{Alesini:2017ifp}
D.~Alesini, D.~Babusci, D.~Di~Gioacchino, C.~Gatti, G.~Lamanna, and C.~Ligi,
  ``{The KLASH Proposal},'' \href{http://arxiv.org/abs/1707.06010}{{\ttfamily
  arXiv:1707.06010 [physics.ins-det]}}.

\bibitem{TheMADMAXWorkingGroup:2016hpc}
{\bfseries MADMAX Working Group} Collaboration, A.~Caldwell, G.~Dvali,
  B.~Majorovits, A.~Millar, G.~Raffelt, J.~Redondo, O.~Reimann, F.~Simon, and
  F.~Steffen, ``{Dielectric Haloscopes: A New Way to Detect Axion Dark
  Matter},'' \href{http://dx.doi.org/10.1103/PhysRevLett.118.091801}{{\em Phys.
  Rev. Lett.} {\bfseries 118} no.~9, (2017) 091801},
  \href{http://arxiv.org/abs/1611.05865}{{\ttfamily arXiv:1611.05865
  [physics.ins-det]}}.

\bibitem{Brun:2019lyf}
{\bfseries MADMAX} Collaboration, P.~Brun {\em et~al.}, ``{A new experimental
  approach to probe QCD axion dark matter in the mass range above 40
  $\mu$eV},'' \href{http://dx.doi.org/10.1140/epjc/s10052-019-6683-x}{{\em Eur.
  Phys. J. C} {\bfseries 79} no.~3, (2019) 186},
  \href{http://arxiv.org/abs/1901.07401}{{\ttfamily arXiv:1901.07401
  [physics.ins-det]}}.

\bibitem{Arvanitaki:2017nhi}
A.~Arvanitaki, S.~Dimopoulos, and K.~Van~Tilburg, ``{Resonant absorption of
  bosonic dark matter in molecules},''
  \href{http://dx.doi.org/10.1103/PhysRevX.8.041001}{{\em Phys. Rev. X}
  {\bfseries 8} no.~4, (2018) 041001},
  \href{http://arxiv.org/abs/1709.05354}{{\ttfamily arXiv:1709.05354
  [hep-ph]}}.

\bibitem{Armengaud:2019uso}
{\bfseries IAXO} Collaboration, E.~Armengaud {\em et~al.}, ``{Physics potential
  of the International Axion Observatory (IAXO)},''
  \href{http://dx.doi.org/10.1088/1475-7516/2019/06/047}{{\em JCAP} {\bfseries
  06} (2019) 047}, \href{http://arxiv.org/abs/1904.09155}{{\ttfamily
  arXiv:1904.09155 [hep-ph]}}.

\bibitem{Aghanim:2018eyx}
{\bfseries Planck} Collaboration, N.~Aghanim {\em et~al.}, ``{Planck 2018
  results. VI. Cosmological parameters},''
  \href{http://dx.doi.org/10.1051/0004-6361/201833910}{{\em Astron. Astrophys.}
  {\bfseries 641} (2020) A6}, \href{http://arxiv.org/abs/1807.06209}{{\ttfamily
  arXiv:1807.06209 [astro-ph.CO]}}.

\bibitem{Abazajian:2016yjj}
{\bfseries CMB-S4} Collaboration, K.~N. Abazajian {\em et~al.}, ``{CMB-S4
  Science Book, First Edition},''
  \href{http://arxiv.org/abs/1610.02743}{{\ttfamily arXiv:1610.02743
  [astro-ph.CO]}}.

\bibitem{Delabrouille:2017rct}
{\bfseries CORE} Collaboration, J.~Delabrouille {\em et~al.}, ``{Exploring
  cosmic origins with CORE: Survey requirements and mission design},''
  \href{http://dx.doi.org/10.1088/1475-7516/2018/04/014}{{\em JCAP} {\bfseries
  04} (2018) 014}, \href{http://arxiv.org/abs/1706.04516}{{\ttfamily
  arXiv:1706.04516 [astro-ph.IM]}}.

\bibitem{Domcke:2020kcp}
V.~Domcke, Y.~Ema, K.~Mukaida, and M.~Yamada, ``{Spontaneous Baryogenesis from
  Axions with Generic Couplings},''
  \href{http://dx.doi.org/10.1007/JHEP08(2020)096}{{\em JHEP} {\bfseries 08}
  (2020) 096}, \href{http://arxiv.org/abs/2006.03148}{{\ttfamily
  arXiv:2006.03148 [hep-ph]}}.

\end{thebibliography}

\providecommand{\href}[2]{#2}\begingroup\raggedright\endgroup

\end{document}